\documentclass{ws-ijmpd}

\usepackage{amsmath}
\usepackage{amssymb}

\begin{document}

\markboth{Garc\'{\i}a \'{A}lvarez} {The logic behind Feynman's
paths}

\title{THE LOGIC BEHIND FEYNMAN'S PATHS\footnote{This work is based on the talks the author gave at \textit{Instituto de
Astronom\'{\i}a y F\'{\i}sica del Espacio}, Buenos Aires, in April
2009, and at \textit{Facultad de Ciencias, Universidad de la
Rep\'{u}blica}, Montevideo, in October 2009.}}

\author{EDGARDO T. GARC\'{I}A \'{A}LVAREZ }

\address{edgardo.physics@gmail.com}

\maketitle

\begin{history}
\received{Day Month Year} \revised{Day Month Year} \comby{Managing
Editor}
\end{history}

\begin{abstract}
The classical notions of continuity and mechanical causality are
left in order to reformulate the Quantum Theory starting from two
principles: I) the intrinsic randomness of quantum process at
microphysical level, II) the projective representations of
symmetries of the system. The second principle determines the
geometry and then a new logic for describing the history of events
(Feynman's paths) that modifies the rules of classical probabilistic
calculus. The notion of classical trajectory is replaced by a
history of spontaneous, random an discontinuous events. So the
theory is reduced to determining the probability distribution for
such histories according with the symmetries of the system. The
representation of the logic in terms of amplitudes leads to Feynman
rules and, alternatively, its representation in terms of projectors
results in the Schwinger trace formula.

\bigskip

\end{abstract}

\section{Introduction}

{\em ...when one does not try to tell which way the electron goes, when
  there is nothing in the experiment to disturb the electrons, then
  one may not say that an electron goes either through hole 1 or hole
  2. If ones does say that, and starts to make any deductions from the
  statement, he will make errors in the analysis.  This is the logical
  tightrope on which we must walk if we wish to describe nature
  successfully.  (R. P. Feynman, Caltech Lectures, 1965).}

\smallskip The purpose of this work is the quest for the first principles
of the Quantum Theory. If we carefully take a look at the axioms of
the mathematical formalisms,\cite{hei30}\cdash\cite{sak94} we will
realize that they are organized by two fundamental ideas: the
projective linear representation of the symmetries of the system and
the intrinsic
aleatory behavior of the events which happen in it.\footnote{%
In the development of Quantum Theory, the concept of chance and
spontaneity of transition, much to our surprise, merges in
Einstein's\cite{ein17} derivation of Planck's formula which finally
crystallizes in Born's\cite{bor35} interpretation of wave function.
However the first antecedent of introducing chance as physical
principle was in kinetic gas theory through Boltzmann's\cite{bol95}
molecular chaos hypothesis.} The ideas mentioned can be summarized
in the following table:
\begin{equation*}
\begin{tabular}{|l|l|}
\hline
\begin{tabular}{l}
\textbf{CHANCE} \\
(Distribution \\
of probabilities)%
\end{tabular}
&
\begin{tabular}{l}
\\
$\bullet $ Born rule \\
$\bullet $ Projection postulate \\
$\bullet $ Sum rule for intermediate events \\
.%
\end{tabular}
\\ \hline
\begin{tabular}{l}
$\text{\textbf{SYMMETRIES}}$ \\
(Projective representation \\
of generators algebra)%
\end{tabular}
&
\begin{tabular}{l}
\\
$\bullet $ Schroedinger equation (Galilei Group) \\
$\bullet $ $\left[ Q,P\right] =i\hslash $ (Heisenberg-Weyl Group) \\
$\bullet $ $\left[ J_{i},J_{j}\right] =i\hslash \epsilon _{ijk}J_{k}$
(Rotation Group) \\
...%
\end{tabular}
\\ \hline
\end{tabular}%
\end{equation*}

In this way, on the one hand, we know that the basic rules, as the
Schroedinger equation and the algebra of generators of the symmetries of the
system, are a consequence of assuming the second descriptive principle
(projective representations of the symmetries of the system).

On the other hand, we shall see that the rules such as the
projection postulate of von Neumann\cite{von32} and Born
rule,\cite{bor35} or alternatively Feynman\cite{fey65b} rules for
combining amplitudes, are a consequence of assuming randomness as
the first principle, in the framework of the second descriptive one.
As we explain in this work, such rules for quantum probabilities are
the democratic way of assigning a distribution of probabilities for
the random transitions between the events of ray space. It means
that all the rays have an equal statistical weight, avoiding any
kind of privileged direction. Therefore, a priori, we will to assume
an isotropic probability distribution in ray space.

Feynman rules, summarized in the next table, represent an equivalent
way of providing the rules for quantum distribution probabilities.
They encode, in terms of amplitudes, the underlying logic of ray
space:\cite{fey65b}

\bigskip

\begin{center}
\begin{tabular}{|l|}
\hline
\ \ \ \ \ \ \ \ \ \ \ \ \ \ \ \ \ \ \ \ \ \ \ \ \ \ \ \ \ \ \ \ \ \ \ \ \
\textbf{FEYNMAN RULES\ } \\
\\
$\bullet $ \textbf{I} -\textbf{Generalization of Born rule:} \textit{the
probability that a particle arrives} \\
\textit{at a }given \textit{position }$x$\textit{, departing from the source
}$s$\textit{, can be represented by} \\
\textit{the square of the absolute value of a complex number called
probability} \\
\textit{amplitude.} \\
$\bullet $ \textbf{II} -\textbf{Sum rule for intermediate alternative events:%
} \textit{when a particle} \\
\textit{can reach a state taking two possible alternative paths, the total}
\\
\textit{amplitude of the process is the sum of the amplitudes of each path}
\\
\textit{considered independently.} \\
$\bullet \mathbf{\ }$\textbf{III-} \textbf{Product rule for consecutive
amplitudes }(which implicitly \\
contains the actualization rule of conditional probabilities): \textit{when
a particle} \\
\textit{follows a path, the amplitude of such path can be written as the
amplitude } \\
\textit{of advancing the first part of the way times the amplitude of
advancing the} \\
\textit{second one.} \\ \hline
\end{tabular}
\end{center}

\bigskip

Generally people think that such rules follow an enigmatic logic, because
they conceive them from the point of view of the classical rules of logic
and probabilities. But, we will see that Feynman rules are not
contradictory, they only encode a natural logic for ray space.

As was recognized by its founding fathers,\cite{hei30} the essential
characteristics of the quantum processes are:

$\bullet $ spontaneity

$\bullet $ randomness

$\bullet $ discontinuity

$\bullet $ bifurcation\qquad

This picture sharply contrasts with the one corresponding to classical
processes, that are inertially stable, deterministic, continuous, and
univocal. In other words, at a quantum level, after any event occurs, it can
spontaneously follow any of the potential events that the interaction
permits. That is, there is no unique history determined by the initial
conditions; precisely because we have a range of possibilities measured by
the probabilities of the aleatory transitions. In this way, Quantum Theory
gives us a physical image of aleatory processes, as if they were
interconnected in a network that takes into account all the possibilities.

Quantum theory often surprises and confuses us since it implies a radical
change of the classical logic. It happens because, in a subtle way, all our
ordinary languages hide epistemological, ontological and logical assumptions
taken from the mechanical conception of the Universe.

In the following section, the prejudices of classical logics will be
discussed and removed. Von Neumann was the pioneer in this
enterprise, realizing that, in order to understand Quantum Theory,
we also must move away from classical logics.

\section{Logic}

In order to understand the logic of quantum processes, first we have
to realize that the theory does not speak about classical
probabilities of sets! Only in the case of Classical Statistical
Mechanics, the probabilities are defined on sets of points of phase
space. Wigner-Moyal representation of the theory perhaps can bring
us this illusion, but it is certainly not the case. Wigner's
distributions are not probabilities.

In order to understand it better, let us briefly review the descriptive
framework of Quantum Theory. Such theory describes the transformations of
symmetries $(T)$ of the systems. As these symmetries have a group structure,
they admit a linear representation $U(T)$ in terms of matrices or linear
operators in a vector space.\footnote{%
Notice that the product of the non-singular square matrix is closed,
associative, has the identity matrix as the neutral element, and inverse
matrix for each element. Therefore, for any group we can find an
homomorphism with square non-singular matrices. Such homomorphism is a
linear representation of the group.} And this is the reason why Quantum
Mechanics works in linear vector spaces (Principle of linear superposition%
\cite{dir58}). In particular, Quantum Theory works with projective
representations in a vector space with a scalar complex
field,\cite{wey31} that are representations up to a phase factor
\begin{equation}
U(T_{1}\circ T_{2})=e^{i\alpha (T_{1},T_{2})}U(T_{1})U(T_{2})
\end{equation}%
due to the fact that all the vectors belonging to the same ray are
physically equivalent.\footnote{%
All the vectors are equivalent up to a complex scale factor, but
working with normalized vectors only the phase is relevant.}
Although phases will be irrelevant in the geometrical examples
discussed in this work, they have
crucial importance\footnote{%
A relative phase of $\pi$ between intrinsic parities of particles
and antiparticles, determines Dirac equation.\cite{gai95}} in the
projective representations of Galilei Group\cite{sch91,sch01} and in
the group of translations in phase space (Heisenberg- Weyl
group.\cite{wey31}) The phases make that the projective
representations of the momentum and the spatial coordinate do not
commute. This is to say that they are the reason for the uncertainty
principle.

So we will focus on the idea that Quantum Theory works on linear
vector spaces. With the sole aim of illustrating it, here we will
consider the simplest non trivial case: The vector space
corresponding to the Euclidean plane. In this case, rotations
$R(\theta )$ are represented by square matrices of the form
\begin{equation}
U\left[ R(\theta )\right] =\left[
\begin{array}{cc}
\cos \theta & \sin \theta \\
-\sin \theta & \cos \theta%
\end{array}%
\right]
\end{equation}%
This matrix rotates vectors in a two dimensional space. However, as
the physical events of the theory $a,b,...$ do not correspond to
vectors $\left\vert a\right\rangle ,\left\vert b\right\rangle ,...$
but to rays or directions in such space, all the vectors are
equivalent up to a scale factor. Then, in some sense, it is not the
Euclidean Geometry who plays the game but Projective Geometry.

Then, with this geometrical picture in mind, notice the following.
If we affirm that the event $a$ ``is" the set of points of ray $a$
and similarly in the case of $b,$ as soon as we try to find the
event corresponding to ``simultaneously being" in $a$ and $b,$ we
obtain the null ray, which has no sense. The problem appears no
sooner than we try to use the set logic theory which only works well
for Classical Mechanics. The words between inverted commas are used
to point out that the problem rests on the fact that the logic of
quantum processes is not combinational but sequential. Quantum logic
is not a logic of states, but one of processes or transitions. As it
was originally pointed out by Bohr and Heisenberg,\cite{hei30,hei72}
the theory does not describe ``states" but processes or transitions
between them.\cite{bau03,fin08} Bear in mind that Quantum Theory was
developed for describing quantum jumps between energy levels. With
these ideas, we have to reformulate the problem. So if we imagine
that the two rays represent a process like an aleatory jump of the
event $a$ to $b,$ the picture recovers its sense.

\bigskip Returning to the geometry of Euclidean plane, notice that the
successive projections of a ray $a$ onto ray $b,$ forming a relative angle $%
\theta _{ab},$ and again onto $a$ result\footnote{%
For example in the base that $P_{a}=\left[
\begin{array}{cc}
1 & 0 \\
0 & 0%
\end{array}%
\right] $ is diagonal$,$ the projector $P_{b}=U(\theta )P_{a}U(-\theta )=%
\left[
\begin{array}{cc}
\cos ^{2}\theta & \cos \theta \sin \theta \\
\cos \theta \sin \theta & \sin ^{2}\theta%
\end{array}%
\right] $ , so $P_{a}P_{b}P_{a}=\cos ^{2}\theta \allowbreak \left[
\begin{array}{cc}
1 & 0 \\
0 & 0%
\end{array}%
\right] ,$ $\theta =\theta _{ab}.$}
\begin{equation}
P_{a}P_{b}P_{a}=\cos ^{2}\theta _{ab}P_{a}
\end{equation}%
because the operation $P_{a}P_{b}P_{a}$ has to be proportional to $P_{a},$ $%
P_{a}P_{b}P_{a}=\lambda P_{a}$ (this is a projection composed with a
dilatation) with $\lambda =\cos ^{2}\theta _{ab}$ . Notice that the
contraction factor $\lambda $ defines a geometrical probability which
coincides with the Born's one. This is a general property of vector spaces
which can be easily verified using Dirac's bra-ket notation ($%
P_{a}=\left\vert a\right\rangle \left\langle a\right\vert $)
\begin{equation}
P_{a}P_{b}P_{a}=\left\langle a\mid b\right\rangle \left\langle b\mid
a\right\rangle P_{a}
\end{equation}

Von Neumann\cite{von32} was the first in associating certainties
with projection operators in his bi-valued logic. According to him,
the projector's eigenvalues $1$ and $0$ of a projector $P_{a}$
indicate if the system ``is" in a given ray $a$ or in the orthogonal
one $\overline{a}$. So the projector $P_{a}$ itself represents a
logical proposition of sharp true values $0$ and $1$. Likewise, the
projector onto the orthogonal complement $\overline{P_{a}}=1-P_{a}$
represents the negation of the proposition $P_{a}.$ It is a nice
evocative idea but with an ontology still contaminated by Classical
Mechanics. We will try to reformulate it in the following sense: the
eigenvalues $1$ and $0$ indicate that we have the certainty of
making a transition to the same ray, but rule out the possibility of
doing it to the orthogonal one. In general, the
projection of a ray $b$ onto $a$ gives as a result the operator $%
P_{a}P_{b}P_{a}$ of eigenvalues $\lambda $ and $0,$ being the first
one a number (between $0$ and $1)$ which measures the degree of
certainty of making the transition. In this way, sharp true values
$0$ and $1$ of von
Neumann's idea are extended to the real number interval $\left[ 0,1%
\right]$.

Summarizing, as we have explained above, the operation
$P_{a}P_{b}P_{a},$ that represents the projection of ray $b$ onto
$a,$ keeps close analogy with the intersection of sets ($\cap $)
which we will denote as
\begin{equation}
P_{b}\sqcap P_{a}=P_{a}P_{b}P_{a}
\end{equation}%
This will be our definition of quantum ``intersection" or AND $(\sqcap )$.
However, it is important to remember that projections in general do not
commute. So our AND between rays is, in general, non commutative. And, this
is precisely the essential distinctive character of the logics for the
quantum processes proposed here, no taken into account by earlier attempts.

If we have a history $\gamma $ in which three consecutive events
$a$, $b$, and $c$ appear in this order, then it represents a logical
proposition that can be rewritten as
\begin{equation}
\gamma =\left( a\sqcap b\right) \sqcap c
\end{equation}

In terms of Feynman rules for combining amplitudes, the total
amplitude of the history can be obtained by multiplying the partial
amplitudes of the chain (in Dirac's notation)
\begin{equation}
A(\gamma )=\left\langle a\mid b\right\rangle \left\langle b\mid
c\right\rangle
\end{equation}%
where we have followed the opposite of Feynman's convention, who prefers to
draw the amplitudes writing the initial event on the right.

If we have a path which bifurcates into two alternative (mutually
excluded) intermediate events, according to Feynman we must add the
amplitudes. This defines the logical operation XOR $(\sqcup )$,
which in terms of projectors reads
\begin{equation}
P_{a}\sqcup P_{b}=P_{a}+P_{b}
\end{equation}

In the following table we summarize the basic operations of the
logic of quantum processes, alternatively represented in terms of
operators and amplitudes:\cite{gar03}
\bigskip
\begin{equation*}
\begin{tabular}{|c|c|c|}
\hline
\textbf{History} & \textbf{Operator} & \textbf{Amplitude} \\ \hline
$\ \gamma $ & $\Gamma (\gamma )$ & $A(\gamma )$ \\ \hline
$a\sqcap b$ & $P_{b}P_{a}P_{b}$ & $\left\langle a\mid b\right\rangle $ \\
\hline
$a\sqcup b$ & $P_{a}+P_{b}$ & ----- \\ \hline
$(a\sqcap b)\sqcap c$ & $P_{c}P_{b}P_{a}P_{b}P_{c}$ & $\left\langle a\mid
b\right\rangle \left\langle b\mid c\right\rangle $ \\ \hline
$(a_{1}\sqcup a_{2})\sqcap b$ & $P_{b}(P_{a_{1}}+P_{a_{2}})P_{b}$ & $%
\left\langle a_{1}\mid b\right\rangle +\left\langle a_{2}\mid b\right\rangle
$ \\ \hline
$\left[ a\sqcap (b_{1}\sqcup b_{2})\right] \sqcap c$ & $%
P_{c}(P_{b_{1}}+P_{b_{2}})P_{a}(P_{b_{1}}+P_{b_{2}})P_{c}$ & $\left\langle
a\mid b_{1}\right\rangle \left\langle b_{1}\mid c\right\rangle $ \\
&  & $+\left\langle a\mid b_{2}\right\rangle \left\langle b_{2}\mid
c\right\rangle $ \\ \hline
\end{tabular}%
\end{equation*}

\bigskip

Any history $\gamma $ has associated a degree of certainty $\lambda (\gamma
) $ (the non trivial eigenvalue of $\Gamma (\gamma )$), that, in terms of
projectors, can be written as\footnote{%
For example, consider the history $\Gamma =\left( P_{a}\sqcap P_{b}\right)
...\sqcap P_{z}$ that starts at $a$ and finishes at $z,$ the non trivial
eigenvalue of $\Gamma $ satisfies $\Gamma P_{z}=\lambda P_{z}$; then taking
the trace in both members we have $\lambda (\gamma )=tr(\Gamma
P_{z})=tr(\Gamma ),$ because $tr(P_{z})=1$.} $tr(\Gamma )$ and, in terms of
amplitudes, as $A(\gamma )A(\gamma ^{-1}),$ where $\gamma ^{-1}$ is the
inverse path of $\gamma $. The equivalent of the representation of the
second and third columns is guaranteed by\footnote{%
For example, in the case of the third row, it is easy to verify that $%
tr\left( P_{c}P_{b}P_{a}P_{b}P_{c}\right) =\left\langle c\mid
b\right\rangle \left\langle b\mid a\right\rangle \left\langle a\mid
b\right\rangle \left\langle b\mid c\right\rangle $ so, we have
$tr\left[ \left( P_{a}\sqcap P_{b}\right) \sqcap P_{c}\right]
=A(a\sqcap b\sqcap c)A(c\sqcap b\sqcap a).$}
\begin{equation}
tr(\Gamma )=A(\gamma )A(\gamma ^{-1})
\end{equation}

Summarizing, quantum histories admit two representations that are
mathematically equivalent: Feynman's\cite{fey65b} representation in
terms of amplitudes which has its roots in the path integral
formalism\cite{fey42}\cdash\cite{fey65c} and the representations in
terms of projectors whose first ideas can be found in the works by
Schwinger on the algebra of measurements.\cite{sch59,sch91,sch01}
Von Neumann\cite{von32} wrote a similar expression in his
\textit{Mathematical Foundations of Quantum Mechanics}, for the case
of commuting projectors. However, neither Feynman nor Schwinger
realized of the logic behind quantum processes. Even though von
Neumann became conscious that the problem lays on the logic, he
finally failed in finding the right one. He did not succeed, because
the quantum logic is not a black and white's one, is a logic of
greys.

The third row of the table represents the history of a polarization
experiment, and the fifth one the history corresponding to the
double slit experiment. These are the two basic experiments chosen
by Dirac\cite{dir58} for introducing the theory in his
\textit{Principles of Quantum Mechanics}. Feynman believed that
these experiments encapsulate all the mystery of the
Quantum.\cite{fey65b,fey85}

The paradox that brings to light the polarization experiment is the
following: suppose that, in the history of the third row, $a$ and
$c$ represent the events corresponding to the passage of a photon
through two filters that polarize light in two orthogonal directions
($c=\overline{a}$); and that $b$ is the event associated with the
passage through an intermediate filter which polarizes light in a
direction on the same plane, forming an angle $\theta _{ab}$ with
the one determined by the first polarizer, and an angle $\theta
_{bc}=\pi /2-$ $\theta _{ab}$ with the second one. The corresponding
quantum history
\begin{equation}
\left( a\sqcap b\right) \sqcap \overline{a}
\end{equation}%
has its analog in the Boolean expression (capital letters indicate
sets)
\begin{equation}
(A\cap B)\cap \overline{A}=\emptyset
\end{equation}%
But, from the point of view of the classical reasoning, the light
would not pass through the third polarizer. But, in spite of the
fact that the projection of ray $a$ onto the orthogonal ray
$\overline{a}$ is null, the magic of quantum logic rests on the fact
that the projection of $a$ onto $b$ and then onto $\overline{a}$ is
different from zero. So, the history has a degree of certainty
\begin{equation}
\lambda =\cos ^{2}(\theta _{ab})\sin ^{2}(\theta _{ab})
\end{equation}%
The explanation is analogous to the tunnel effect. A classically forbidden
history, as the passage through a potential barrier, is only possible at
quantum level because intermediate events that have non null projections.

In the case of the double slits experiment, the mystery is the
generation of an interference pattern. But, if we follow the logic
of Quantum processes, the interference is an unavoidable
consequence. In fact, the operator corresponding to the history in
which the electron can pass ``alternatively"\footnote{It is
important not to fall in the trap of ordinary language, that
interprets ``alternatively" in Boolean sense. Here ``alternatively"
must be read in terms of the connective $\sqcup .$ In other words,
from the proposition $\left[ a\sqcap (b_{1}\sqcup b_{2})\right]
\sqcap c$ (the electron passed by $(b_{1}\sqcup b_{2})$) it does not
follows that it passed by the slit $b_{1}$ or $(\sqcup )$ \ by the
slit $b_{2}$ because the
interference term $I$ is different from zero.} by the two slits (event $%
b_{1}\sqcup b_{2}$) is the the sum of the operators representing the
histories in which the electron effectively passes by each slit plus
an interference term $(I):$
\begin{equation}
\Gamma _{\left[ a\sqcap (b_{1}\sqcup b_{2})\right] \sqcap c}=\left(
P_{a}\sqcap P_{b_{1}}\right) \sqcap P_{c}+\left( P_{a}\sqcap
P_{b_{2}}\right) \sqcap P_{c}+I
\end{equation}%
with
\begin{equation}
I=P_{c}P_{b_{1}}P_{a}P_{b_{2}}P_{c}+P_{c}P_{b_{2}}P_{a}P_{b_{1}}P_{c}
\end{equation}%
Again the paradox only appears when we insist in reasoning
classically. In this case, the analogous Boolean expression would be
\begin{equation}
A\cap (B_{1}\cup B_{2})\cap C=\left( A\cap B_{1}\cap C\right) \cup \left(
A\cap B_{2}\cap C\right)
\end{equation}%
which leads to think in terms of Kolmogorov's additive probabilities
for disjoint events.\cite{kol56}

The crossed terms between projectors $P_{b_{1}}P_{a}P_{b_{2}}
+P_{b_{2}}P_{a}P_{b_{1}}$, associated to the orthogonal rays $b_{1}$ and $%
b_{2}$, are responsible for the characteristic interference of the
quantum phenomena. For instance, crossed terms between positive
(electron) and negative energy levels (positron) are responsible for
the trembling motion of the positron-electron charge
(\textit{Zitterbewegung}) which, in each subspace of definite sign,
acquires a magnetic moment.\footnote{See Ref.~\refcite{gai98} and
references cited therein.} It is immediate to see that, if $P_{a}$
commutes with $P_{b_{1}}$ and $P_{b_{2}},$ then the interference
vanishes. But, again, it is a property of vectorial spaces that the
successive action of three projectors such as
$P_{b_{1}}P_{a}P_{b_{2}}$ can be different from zero. In other
words, the mystery of the quantum behavior rests on the logic of
projections, since the event space is the ray space. In other words,
the slits do not play as filters in ordinary space but in ray space.

For those who feel the vertigo of \textit{making equilibrium over
the logic tightrope of Quantum Theory},\cite{fey65b} the amplitude
representation offers a momentary relaxation that maintains some
parts of the reasoning in terms of Boolean logic. In this case, it
is enough to follow Feynman rules II and III for computing the total
amplitude of the path $\gamma =\left[ a\sqcap (b_{1}\sqcup
b_{2})\right] \sqcap c$ . For counting all possible paths and
calculating partial and total amplitudes, only ordinary logic is
needed. But the calm is just temporary, because at the end of the
day, for getting the probabilities for amplitudes, we have to use
rule I, and multiply this amplitude by the amplitude of the reversed
path $\gamma =\left[ c\sqcap (b_{1}\sqcup b_{2})\right] \sqcap a$.
In general, as can be easily verified, this procedure is equivalent
to taking into account all the closed paths that start from the
initial event and come back. In fact, in our
example, it is easy to check that\footnote{%
When in order to abbreviate, we omit the parenthesis, it is assumed the
operation $\sqcap $ is taken in sequential order. At this point is
instructive to calculate probability in both representations. That is taking
the trace of $\Gamma $ and squaring the absolute value of the amplitude $%
A(\gamma )$.}
\begin{equation}
tr(\Gamma )=A(a\sqcap b_{1}\sqcap c\sqcap b_{1}\sqcap a)+A(a\sqcap
b_{2}\sqcap c\sqcap b_{2}\sqcap a)+tr(I)
\end{equation}%
where
\begin{equation}
tr(I)=A(a\sqcap b_{1}\sqcap c\sqcap b_{2}\sqcap a)+A(a\sqcap b_{2}\sqcap
c\sqcap b_{1}\sqcap a).
\end{equation}%
Paying attention to the last expressions, it immediately follows
that we not only have to compute the closed path which comes back by
taking the same path in the opposite sense, but we have also to take
into account the closed loops that go through one slit and come back
through the other one. These two paths $\gamma _{\circlearrowleft
}=a\sqcap b_{1}\sqcap c\sqcap b_{2}\sqcap a$ and $\gamma
_{\circlearrowright }=a\sqcap b_{2}\sqcap c\sqcap b_{1}\sqcap a$,
that circulate in reverse sense, are responsible for the
interference
\begin{equation}
tr(I)=A(\gamma _{\circlearrowleft })+A(\gamma _{\circlearrowright })
\end{equation}%
In other words, as the probability of a path is the amplitude of going
forward and coming backward, this simple rule for the ``logic of paths" open
the possibility of having closed loops which enclose ``area" different from
zero.\footnote{%
In general there is no area in the ordinary sense (the space can be
discrete), unless the ray space has a continuum spectrum as is the
case of momentum and coordinate representations.} This is a general
property and, perhaps, the most striking example is the
Aharonov-Bohm\cite{aha59} effect, in which the amplitude of the
closed loop which encircles the solenoid acquires a phase
proportional to the magnetic flux. But, probably, the most relevant
one is the path integral in phase space itself. In this case, closed
paths enclose an area
\begin{equation}
S(\gamma _{\circlearrowleft })=\oint p\, dx-Hdt
\end{equation}%
in an extended phase space of canonical variables $(p,H)$ and $(x,t),$ which
represent the action of these paths.\footnote{%
In general $H$ is not the classical Hamiltonian, but the Wigner's
equivalent of the corresponding operator.\cite{gar93}} This area in
units of $\hslash $ $\left( \frac{S}{\hslash }\right) $ is the phase
of the closed amplitude which contributes to the interference terms.
In the classical regime we have big phases which highly oscillate;
thus, they destructively interfere. The main contribution to the
total probability comes from the close surroundings of the path,
whose phase is stationary ($\delta S=0),$ that is to say, the path
corresponding to the classical trajectory derived from the principle
of least action.\cite{fey42,fey48,fey65c,fey65a} \footnote{See also
Ref.~\refcite{fey85} for a nice derivation of geometrical optics
starting
from Feynman rules.} In this way, the choice among all the possibilities%
\textbf{\ }in the network of the whole potential events (the path integral
picture) is\textbf{\ }taken by the ``laws" of chance,\footnote{%
The law of big numbers for quantum probabilities was proved by
Finkelstein,\cite{fin96} departing form Born Rule.} recovering the
determinism at a classical scale. \textit{\ }

\section{Probability}

To conclude, we desire to emphasize that Born rule in Quantum Theory
plays the same role as the Pythagoras' theorem in Euclidean
geometry. In a framework such as Analytic Geometry, one can decide
to take it as starting point to introduce the notion of distance.
Otherwise, one looks for another framework based on basic
assumptions, in order to derive it as a theorem, as is the case of
the synthetic Euclidean Geometry. This last point of view was
adopted by the author in Ref.~\refcite{gar03} where the logic of
quantum processes was taken as the basic assumption. For this
purpose, it is necessary to generalize Kolmogorov's
axioms,\cite{kol56} originally developed for the classical theory of
sets. Here, using Laplace's notion of probability, we show in a more
heuristic way, how its rule can be derived. First, we will argue
that, in general, the absolute probability $p(\gamma )$ of a history
is given by:

\begin{equation*}
\begin{tabular}{|c|c|c|}
\hline \textbf{Probability} & \textbf{Operator representation} &
\textbf{Amplitude representation} \\ \hline $p(\gamma /I)$ &
$\frac{1}{N}tr(\Gamma )$ & $\frac{1}{N}A(\gamma )A(\gamma ^{-1})$ \\
\hline
\end{tabular}%
\bigskip\end{equation*}%
where $N=trI$ is the dimension of the space ray, and $I$, the identity
matrix, represents the universe of sample space.

For turning the ideas more concrete, we will consider the idealized model of
a \textit{quantum} die, essentially a six-level system. A quantum die
differs from a classical one in the sense that all its faces represent
orthogonal rays in ray space, because the corresponding events are mutually
exclusive.

We are going to assume that when playing with Einstein, \textit{the
Lord is subtle but not malicious}, so the die is ``perfectly
balanced"; therefore all faces have the same a priori probability
(isotropy of ray space). In this case, Laplace would say that the
probability of obtaining any face is $1/6$ (number of cases divided
the number of total possible ones). The reasoning can be formalized
as follows: the event $A$, e.g. obtaining the face $5$, is given by
the set $A=\left\{ 5\right\} ,$ a subset of the sample space
$E=\left\{ 1,2,3,4,5,6\right\}$. So, under equal \textit{a priori}
probabilities for any elements of the sample space, the probability
of obtaining $A$ is directly calculated as the ratio of the cardinal
of set $A$ ($card(A)$) to the cardinal of the sample space
\begin{equation}
p(A/E)=\frac{card(A)}{card(E)}=\frac{1}{6}
\end{equation}%
Similarly, in the quantum case, events are represented by projectors, for
instance, the face $f_{5}$ is represented by the operator $P_{f5,}$ and the
sample space by the development of the identity $%
I=P_{f_{1}}+P_{f_{2}}+P_{f_{3}}+P_{f_{4}}+P_{f_{5}}+P_{f_{6}}.$ But,
in order to obtain a probability equal to $\frac{1}{6}$, we cannot
take cardinals because projectors are not sets! Therefore, we have
to generalize this notion in ray space. It is easy to convince
oneself that the analogous operation is taking the trace
\begin{equation}
p(P_{f_{5}}/I)=\frac{tr(P_{f_{5}})}{tr(I)}=\frac{1}{6}
\end{equation}%
In fact, tracing is the natural generalization of counting. For example:
each elementary projector has a unit trace and the trace of the identity is
the dimension of the space. Moreover, as the trace is invariant under
rotations, so it is the only invariant counting we can define in ray space.

However, in contrast with the classical dice, quantum dice admit more than
one representation of the sample space (``Bohr's complementary principle").
We can rotate the die and obtain a new base of orthogonal faces represented
by the projectors $%
P_{R(f_{1})}=U(R)P_{f_{2}}U^{-1}(R),P_{R(f_{2})},...P_{R(f_{6})}.$ Then we
can ask the following question: if we know, for example, that the face $%
a=f_{2}$ has gone out, which is the conditional probability of
obtaining the face $b=R(f_{2})$? In this case, this is the history
$\gamma =a\sqcap b,$ which is represented by the operator $\Gamma
(\gamma )=P_{b}P_{a}P_{b}.$ But, as we have seen
\begin{equation}
P_{b}P_{a}P_{b}=\cos ^{2}\theta _{ab}P_{b}.
\end{equation}%
So tracing both sides of the equality
\begin{equation}
\frac{tr\left( P_{b}P_{a}P_{b}\right) }{trP_{b}}=\cos ^{2}\theta _{ab}
\end{equation}%
or equivalently
\begin{equation}
\frac{\frac{1}{6}tr\left( P_{b}P_{a}P_{b}\right) }{\frac{1}{6}trP_{a}}=\cos
^{2}\theta _{ab}
\end{equation}%
since the trace of any elementary projector is one.

The first member of the equality is the conditional probability of
obtaining the face $b,$ having obtained the face $a$ before,
\begin{equation}
p(a\sqcap b/a)=\frac{p(a\sqcap b)}{p(a)}=\cos ^{2}\theta _{ab},
\end{equation}%
which coincides with the expression originally proposed by Born.$\square $

In general, the probability of a history $\gamma $ which begins at an
initial event $a$, results in
\begin{equation}
p(\gamma /a)=\frac{p(\gamma \sqcap a)}{p(a)}=tr(\Gamma )
\end{equation}%
which coincides with the degree of certainty $\lambda (\gamma )$ of
the history. The right side of last expression was originally
proposed by Schwinger\cite{sch59} in the context of his algebra of
measurements.\cite{sch91,sch01} Later, the trace formula was
rediscovered, but consistent conditions were imposed \textit{ad hoc}
in order to keep Boolean logic. However, the only logic that
reproduces the standard formalism of Quantum Theory is the one
exposed in this paper.

Let us consider again the history of three events $\lambda =a\sqcap
b\sqcap c.$ Then, if we compute the conditional probability of the
last event $c$ given the two first $a\sqcap b$ we have
\begin{eqnarray}
p(c/a\sqcap b) &=&\frac{p\left[ \left( a\sqcap b\right) \sqcap c\right] }{%
p(a\sqcap b)} \notag \\ \\
&=&\frac{tr(P_{c}P_{b}P_{a}P_{b}P_{c})}{tr(P_{b}P_{a}P_{b})}
\end{eqnarray}%
Now, if we look this expression carefully, we observe it can be
rewritten as
\begin{equation}
p(c/a\sqcap b)=tr(P_{c}P_{\Psi }P_{c})
\end{equation}%
where the operator
\begin{equation}
P_{\Psi }=\frac{P_{b}P_{a}P_{b}}{tr(P_{b}P_{a}P_{b})}
\end{equation}%
is equivalent to the projection operator onto ray $b$. In this way,
the expression is reduced to the one corresponding to the
conditional probability for the history of two events $\Psi \sqcap
c,$ that is
\begin{equation}
p(c/a\sqcap b)=p(c/\Psi )
\end{equation}%
and, as $\Psi =b,$ the last expression indicates us that: after that event $%
b $ has happened, the probability of the original history does not
remember the event $a.\square $ All works as if the history, after
projection onto $b$, begins in this event. This actualization
property of conditional probabilities is equivalent to the von
Neumann rule.

\section{Final remarks and conclusions}

Physics has inherited the notions of Classical Mechanics. The physical
explanation of the problem of motion is essentially deterministic. So, we
have seen that is necessary to dig deeply in order to find the roots of the
problem. And, above all, it is imperative to cut these roots in order to
understand Quantum Theory. This is not an easy enterprise due to the fact
that mechanic philosophy is ubiquitous in our language and has installed in
our minds. The most difficult point to interpret is understanding the reason
of movement. Notice that, contrary to what happens in the mechanical
Universe, at quantum scale, the interactions do not cause changes; they only
open the door so as the transitions can take place. The quantum processes
occur spontaneously by chance.

The interpretations of Quantum Theory that try to conserve the ontology of
Classical Mechanics exacerbate the role of Schroedinger's equation in the
theory. It is claimed that the state of the system evolves continuously, as
if after an event the system followed by ``inertia" the continuous evolution
of probabilities determined by this equation.\footnote{%
Sometimes people confuse the symmetry of the temporal evolution with
the actual history of the system. However, the so called ``temporal
evolution of the system" is not the history of the events that
really happened but a probabilistic description of the potential
events that could have happened.} But this path takes us to a dead
end: explaining the discontinuity through continuity, and chance by
determinism.

So, the conceptual revolution of Quantum Theory will only be completed after
the new conception of the problem of motion will be generally accepted. On
the whole, quantum phenomena show us that aleatory processes are spontaneous
and chance plays the role of the ``inertia principle" in the theory (the
spontaneous persistence of motion). Any other attempt would be another way
of returning to the theory of ``hidden variables".

At present, the big question is the opposite one. How to explain the
apparent determinism we observe at classical level from the aleatory
behavior quantum phenomena. The first answer to this question was
outlined by Dirac;\cite{dir33,dir58} then Feynman\cite{fey42,fey48}
completed this elegant idea developing his path integral formalism.
In this framework, all paths, continuous or discontinuous, are
allowed and the classical trajectory is just the most probable
sequence in the network of potential events.

In conclusion, the paradigm of a legal Universe is absent in Quantum
Theory. Physical ``laws" are just symmetries or reduced to the ``law
of big numbers". Chance is the true physical principle of the
theory. The second principle, symmetry, only plays the role of a
descriptive one. In fact, it proposes a framework for the theory
from which the logic and probabilistic distributions for the
physical processes can be deduced.\footnote{ The suggestion that
there may not be any fundamental dynamical laws was also made by
Anandan.\cite{ana99} He also claimed that the non existence of laws
imply that there can be neither deterministic laws nor fundamental
probabilistic ones. So, transition probabilities can be just
determined by symmetries. Therefore Born rule has not any
fundamental status. } In fact, as the processes are random, we are
just limited (as in the case of the die) to find the symmetries in
the system which allow us to determine the probability for these
processes.

As symmetries have the group mathematical structure, we can establish an
homomorphism with the non singular square matrix. That is a linear
representation of the group of symmetries of the system. In particular, the
mathematical framework of Quantum Theory is the projective representation in
a complex ray space.\footnote{%
The non trivial point is if we are limited to use complex numbers as the
scalar fields. In fact instead of having a group $U(1)$ for the phases we
can generalize it to another internal group of symmetry.} This is to say
that the representation of rays and matrices are defined up to a phase
factor, because, at the end of the day, probabilities only depend on the
square of the amplitudes. This phase factor is a non trivial element. This
one is responsible for the non commutativity of translations in phase space
and, in general, for the interference effects characteristic of quantum
processes. Moreover, it also explains the classical limit of the theory.

To sum up, we have seen that the geometry of the ray space
determines the logic of the quantum processes and, as a consequence,
the quantum probabilistic rules such as Born's and von Neumann's
ones, historically postulated \textit{ad hoc}. This logic and
probabilistic rules a priori sound enigmatic, since we have to leave
aside the familiar Boolean logic and Kolmogorov's notion of
probabilities. However, if we finally accept that the ray space is
the true scenery where the events occur, the new logic is inherently
derived from geometry. The propositions of this logic are chains of
sequential paths of events (histories) that have two equivalent
representations. One uses of projector operators, and the other one
works in terms of complex amplitudes which follow Feynman rules. In
this rules, which describe the basic experiments of polarization and
interference, is encapsulated the mysterious logic of the quantum
processes.

\section*{Acknowledgments}
I am in debt to Fabian Gaioli, David Finkelstein, Larry Horwitz for
nice discussions and encouragement. Also to Rodolfo Gambini and
Jorge Pullin for their kind hospitality at Universidad de la
Rep\'{u}blica (Montevideo) where I gave a second talk on this work.
Finally, I want to thank Mario Castagnino for inviting me to give
the first talk on this work, and for all the enthusiasm he still
transmits at his young 75 years.


\begin{thebibliography}{99}
\bibitem{hei30} W. Heisenberg, \textit{The physical principles of Quantum
Theory }(1930), reprinted by (Dover, New York, 1949).

\bibitem{hei72} W. Heisenberg, \textit{Development of concepts in the
History of Quantum Mechanics }(1972), reprinted in \textit{Encounters with
Einstein, and other essays on People, Places and Particles} (Princeton
University Press, Princeton, 1989).

\bibitem{dir58} P. A. M. Dirac, \textit{The Principles of Quantum Mechanics},
(Oxford University Press, Oxford, 1958).

\bibitem{mes83} A. Messiah, \textit{Mec\'{a}nica Cu\'{a}ntica}, (Tecnos,
Madrid, 1983).

\bibitem{coh77} C. Cohen-Tannoudji, B. Diu and F. Laloe, \textit{Quantum
Mechanics,} (Wiley, New York,1977).

\bibitem{sak94} J. J. Sakurai, \textit{Modern Quantum Mechanics}, (Adison-Wesley,
Massachusetts, 1994).

\bibitem{ein17} A. Einstein, \textit{On the Quantum Theory of Radiation},
Phys. Z. \textbf{18} (1917) 121, reprinted in \textit{Source of Quantum
Mechanics}, edited by Van der Waerden B. L., (Dover, New York, 1968).

\bibitem{bor35} M. Born, \textit{Atomic Physics}, (1935) reprinted by
(Dover, New York, 1989).

\bibitem{bol95} L. Boltzmann, \textit{Lectures on Gas theory }(1895),
reprinted by (Dover, New York, 1995).

\bibitem{von32} J. von Neumann, \textit{Mathematical Foundations of Quantum
Mechanics }(1932), reprinted by Consejo Superior de Investigaciones
Cient\'{\i}ficas, (Raycar, Madrid, 1991).

\bibitem{fey65b} R. P. Feynman, R. B. Leighton and M. Sands, \textit{The
Feynman Lectures on Physics Volume III},(Adisson Wesley,
Massachusetts, 1965).

\bibitem{wey31} H. Weyl, \textit{The Theory of Groups and Quantum Mechanics }%
(1931), reprinted in by (Dover, New York, 1950).

\bibitem{gai95} F. H. Gaioli and E. T. Garc\'{\i}a \'{A}lvarez, \textit{%
Am. J. Phys.} \textbf{63} (1995) 177. hep-th/9807211.

\bibitem{sch91} J. Schwinger, \textit{Quantum Kinmetic an Dynamics},
(Addison Wesley, New York, 1991).

\bibitem{sch01} J. Schwinger, \textit{Quantum Mechanics, Symbolism of
Atomics Measurements}, (Springer, Berlin, 2001).

\bibitem{gar03} E. T. Garc\'{\i}a \'{A}lvarez, \textit{Projective Geometry,
Logic and Probability in Quantum Theory }(2003), unpublished.

\bibitem{bau03} D. R. Finkelstein, \textit{Int. J. Theor. Phys.} \textbf{42} (2003) 177.
See also  J. Baugh, D. R. Finkelstein and A. Galiautdinov, hep-th/0206036.

\bibitem{fin08} D. R. Finkelstein, private communication (2008).

\bibitem{fey42} R. P. Feynman, \textit{The Principle of Least Action in
Quantum Mechanics}, Feynman's Thesis (1942), reprinted by L. M.
Brown in \textit{Feynman's Thesis, a new approach to Quantum
Theory}, (World Scientific, Singapore, 2005).

\bibitem{fey48} R. P. Feynman. \textit{Rev. Mod. Phys. }\textbf{20}
(1948) 367.

\bibitem{fey65c} R. P. Feynman and A. R. Hibbs, \textit{Quantum Mechanics
and Path Integrals}, (McGraw-Hill, New York, 1965).

\bibitem{sch59} J. Schwinger, \textit{Proc. Nat. Acad.
Sci.} \textbf{45} (1959) 1552.

\bibitem{fey85} R. P. Feynman, \textit{QED The Strange Theory of Light and
Matter}, (Princeton University Press, Princeton, 1985).

\bibitem{kol56} A. N. Kolmogorov,\textit{The theory of Probability} (1956)
in \textit{Mathematics its Content, Methods and Meaning, }eds. A. D.
Aleksandrov, A. N. Kolmogorov and M. A. Lavrentev, reprinted in
(Dover, New York, 1999).

\bibitem{gai98} F. H. Gaioli and E. T. Garc\'{\i}a \'{A}lvarez, \textit{%
Found. Phys.} \textbf{28} (1998) 1539. hep-th/98731.

\bibitem{aha59} Y. Aharonov and D. Bohm, \textit{Phys. Rev.} \textbf{%
115} (1959) 485.

\bibitem{gar93} E. T. Garc\'{\i}a \'{A}lvarez, Anales de la Asociaci\'{o}n F%
\'{\i}sica Argentina, \textbf{5 }(1993) 9.

\bibitem{fey65a} R. P. Feynman, R. B. Leighton and M. Sands, \textit{The
Feynman Lectures on Physics Volume II, }(Adisson Wesley, Massachusetts,
1965), Chap. 19.

\bibitem{fin96} D. R. Finkelstein, \textit{Quantum Relativity, A synthesis
of the ideas of Einstein and Heisenberg}, (Springer, Berlin, 1996), Chap. 8.

\bibitem{dir33} P. M. A. Dirac, \textit{Physikalische Zeitschrift der
Sowjetunion}, Band3, Heft 1, (1933) 64, reprinted in J. Schwinger \textit{%
Selected papers on Quantum Electrodynamics}, (Dover, New York, 1958).

\bibitem{ana99} J. S. Anandan, \textit{Found. Phys.} \textbf{29}
(1999) 1647. quant-ph/9808045.\bigskip
\end{thebibliography}
\end{document}